\documentclass[12pt]{article}

\newcommand{\pmSig}{\,^{\pm}\!\Sigma}
\newcommand{\pSig}{\,^+\!\Sigma}
\newcommand{\mSig}{\,^-\!\Sigma}
\newcommand{\pbpi}{\,^+\!\mbox{\boldmath$\pi$}}
\newcommand{\mbpi}{\,^-\!\mbox{\boldmath$\pi$}}
\newcommand{\bpi}{\mbox{\boldmath$\pi$}}
\newcommand{\bl}{\mbox{\boldmath$l$}}
\renewcommand{\d}{{\rm d}}
\newcommand{\ppi}{\,^+\!\pi}
\newcommand{\mpi}{\,^-\!\pi}

\begin{document}

\title{On the area expectation values in area tensor Regge calculus in the Lorentzian
domain}
\author{V.M. Khatsymovsky \\
 {\em Budker Institute of Nuclear Physics} \\ {\em
 Novosibirsk,
 630090,
 Russia}
\\ {\em E-mail address: khatsym@inp.nsk.su}}
\date{}
\maketitle
\begin{abstract}
Wick rotation in area tensor Regge calculus is considered. The heuristical expectation
is confirmed that the Lorentzian quantum measure on a spacelike area should coincide
with the Euclidean measu\-re at the same argument. The consequence is validity of
probabilistic interpretation of the Lorentzian measure as well (on the real, i.e.
spacelike areas).
\end{abstract}
\newpage
The problem of the relation between the Lorentzian and Euclidean versions of a field
theory is especially nontrivial in gravity, namely, in general relativity theory where
the Wick rotation means not only formal rescaling time and timelike vector components
by $\sqrt{-1}$ but also real change of topology \cite{Haw}. The discrete counterpart
of general relativity, Regge calculus \cite{Reg} (RC), provides a simplification
connected with coordinatelessness of this formulation of general relativity: there is
no time coordinate to be scaled by $\sqrt{-1}$; instead, some edge lengths should be
made complex (the notion of "time" can arise at the intermediate steps simply as a
parameter which labels the 3D layers by ..., 1, 2, 3, ...) \cite{Wil}.

In the present paper we consider Wick rotation in the framework of the approach to
quantisation of RC developed in a number of our papers. To begin with, we shortly
discuss that quantization problem. Up to now the canonical (Dirac) quantization
prescription is considered as a fundamental principle not contradicting to experiment,
so let us adhere to it. However, there is a problem to implement it in the discrete
theory such as RC because of the lack of a regular continuous coordinate playing the
role of time. Although quantum theory still can be formulated with the help of the
functional integral in this case, the absence of a strict framework leaves the quantum
measure practically arbitrary. In fact, various measures are used in applications
(e.g., in phase analysis of simplicial quantum gravity \cite{HamWil1,HamWil2}).

Despite of the lack of a continuous coordinate in the completely discrete RC, such the
coordinate is present in the limiting so-called (3+1) RC when one of the coordinates
is made continuous by shrinking sizes of all the simplices along this coordinate to
those infinitely close to zero. Therefore one can try to develop Hamiltonian formalism
and canonical quantization and represent the result in the form of the Feynman path
integral measure. Next it is natural to ask whether a measure in the original
completely discrete RC exists such that one could define the limit of this measure
when one of the coordinates is made continuous and the limiting measure would coincide
with the above Feynman path (canonical quantization) measure with this continuous
coordinate playing the role of time. Equivalence of the different coordinates means
that this situation should take place irrespectively of what coordinate is made
continuous.

A difficulty with RC in the continuous time limit is that the description of the
infinitely flattened in some direction simplex purely in terms of the lengths is
singular. The idea is to use description in terms of the variables of the types of
both lengths and angles. This might be achieved in the Regge analog of the
Hilbert-Palatini form of the Einstein action. The discrete analogs of the tetrad and
connection, link vectors and finite rotation matrices, were first considered by Bander
\cite{Ban1,Ban2,Ban3}. We have found exact representations of the Regge action in
terms of link vectors and finite rotation matrices as {\it independent} variables
\cite{Kha0}. This representation results in the exact Regge action if rotation
matrices are excluded via equations of motion, that is, on classical level. Rotation
matrices are just the desired angle type variables which allow us to formulate
continuous time (3+1) RC in a nonsingular way. After that the above strategy can be
implemented. In \cite{Kha3} we write out canonical form of RC.

Next we can try to solve the problem of finding measure in the full discrete RC which
has the desired continuous time limit corresponding to the canonical quantization
irrespectively of what coordinate is taken as a time and made continuous. Although
this last condition is rather restrictive, the problem has solution in 3 dimensions
\cite{Kha2}. In 4 dimensions solution can be found for a certain version of the
so-called "area RC" where areas are treated as independent variables
\cite{BarRocWil,RegWil}. Since the number of areas is larger than the number of
lengths, this means that the lengths of the same edge defined in the different
simplices are in general different, i.e. ambiguous. The configuration superspace of
the area RC contains the hypersurface corresponding to the ordinary RC; at the same
time it is exactly soluble just as 3D model. Appropriate version in our case is "area
tensor RC" with independent {\it area tensors} (i.e., in particular, in general there
are no link vectors corresponding to them). Just the corresponding superspace
(extended as compared to that of ordinary RC) is that space on which the measure under
above restrictive conditions can be found \cite{Kha}.

Remarkable feature simplifying solution of the above problem of finding full discrete
measure for the 3D RC and for the 4D area tensor RC in the tetrad-connection variables
is commutativity of the constraints (arising in the Hamiltonian formalism, i.e. in the
continuous time limit). These constraints as well as canonical quantization itself are
analogous to their completely continuum counterparts. The commuting constraints for
the 3D discrete gravity were first deduced by Waelbroeck \cite{Wael} for general
system (not a'priori restricted to be RC). Analogous first class system of constraints
arises in the area tensor RC in the above mentioned derivation of the measure
\cite{Kha}.

Finally, we need to reduce the quantum measure in the extended superspace to the RC
hypersurface. The idea is to consider area tensor RC system and ordinary RC system as
particular case of the simplicial complex with discontinuous metrics. The point is
that the piecewise flat manifold possesses metric whose normal component undergoes
discontinuity when passing across any 3D face, but the tangential components remain
unchanged. Now we go further and consider system where tangential components of metric
are also discontinuous. It is the system with independent simplices which do not
necessarily fit each other on their common faces. In the superspace of all the
simplicial discontinuous metrics RC corresponds to the hypersurface singled out by
conditions of the tangential metric continuity on the faces. The problem is to reduce
the above constructed quantum measure in "area tensor RC" to this hypersurface. For
that some $\delta$-function-like factor is introduced in the measure which fixes
equality of tangential metric across any face. This factor is found in our paper
\cite{Kha1} by using the principle of "minimum of the lattice artefacts". Namely, we
require that the factor should not depend on the form and size of any face across
which metrics are compared, only on the hyperplane in which the given face is placed.
We show that such the factor preserving equivalence of the different simplices exists
and is unique. Consequences of the viewpoint on area RC (now not tensor one) as a
system with discontinuous metrics were also discussed in \cite{WaiWil}.

It is quite natural that at the intermediate stage of our construction in area tensor
RC the measure is factorizable (on certain conditions) since area tensors are here
independent just as edge vectors in the (exactly soluble) 3D model, the typical factor
as applied to averaging a function $f(\pi)$ of a given area tensor $\pi$ being of the
form (in the case of the Euclidean spacetime signature)
\begin{eqnarray}                                                               
\label{Euclidean_measure}%
<f(\pi)> & = & \int{f (-i\pi ){\rm d}^6\pi\int{e^{\textstyle i\pi\circ R}{\cal D}R}}.
\end{eqnarray}

\noindent Here rotation of the integration contours used to define integral is
performed via substitution of the integration (dummy) variables $\pi$ $\to$ $-i\pi$.
The integration variable $R$ is SO(4) matrix, $A\circ B$ $\equiv$ $A^{ab}B_{ab}/2$.
The formula (\ref{Euclidean_measure}) looks reasonable from path integral and symmetry
viewpoint: $\pi\circ R$ reminds Regge action in the sense that it (for independent
tensors $\pi^{ab}$) gives the same constraints $R$ = 0 arising in the canonical
formalism; ${\rm d}^6\pi$ and ${\cal D}R$ are the invariant Lebesgue and Haar
measures. Upon splitting antisymmetric matrices into self- and antiselfdual ones like
\begin{equation}                                                               
\pi_{ab} \equiv {1\over 2}\ppi_k\pSig^k_{ab} + {1\over 2}\mpi_k\mSig^k_{ab}
\end{equation}

\noindent (the basis of self- and antiselfdual matrices $i\pmSig^k_{ab}$ obeys the
Pauli matrix algebra) the formula (\ref{Euclidean_measure}) reads
\begin{eqnarray}                                                               
\label{Euclidean_measure2}%
<f(\pi)> & = & \int{f (\pi ) {\nu (|\pbpi |)\over |\pbpi |^2}{\nu (|\mbpi |)\over
|\mbpi |^2} {\d^3\pbpi\over 4\pi} {\d^3\mbpi\over 4\pi}}, \\ & & \nu(l)={l\over\pi}
\int\limits_{0}^{\pi}{{\rm d}\varphi\over\sin^2{\!\varphi}}\,{e^{\textstyle
-l/\sin{\varphi}}}. \nonumber
\end{eqnarray}

\noindent This is positive measure which gives finite (due to exponential cut-off)
nonzero expectation values of positive powers of area (and even of negative powers
$|\pi|^k$ at $k$ $>$ $-2$, $|\pi|$ $\equiv$ $(\pi\circ\pi)^{1/2}$) which thus are
certain numbers in the Plank units. Further, "glueing" together neighbouring
4-simplices on their common faces \cite{Kha1} we can go over from the considered area
tensor RC to the genuine RC and make qualitative scaling estimate showing that the
length expectation values are finite nonzero \cite{Kha4}. The properties of the
measure such as positivity and finite nonzero area (length) expectation values remain
true.

Also note the following. If the matrices $R$ in the formula (\ref{Euclidean_measure})
were substituted by their continuum analogs, antisymmetric matrices (elements of the
Lie group so(4)), we would have $\delta$-function-like measure and {\it zero}
expectation values for nonconstant monomials in $\pi^{ab}$. Thus, nonzero area
expectation values arise due to nonlinearity of the {\it finite} rotations, i. e.
eventually are connected with the discreteness of the theory.

Nonvanishing and finiteness of the area (length) expectation values may speak well for
the internal consistency on dynamics level of using the discrete type variables to
describe gravity. If these expectations were equal to zero, this would mean that the
functional integral is saturated by smooth manifolds which are just the limiting case
of the piecewise flat manifolds if edge lengths tend to zero; that is, we would return
to the continuum theory.

More detailed and still concise discussion of the above points is given in our work
\cite{Kha5}.

When passing from Euclidean to the Lorentzian domain the Euclidean area becomes
spacelike (real) Lorentzian area, and one would expect that positivity of the measure
in the Euclidean domain would correspond to positivity in the Lorentzian domain on
real areas. The technical complication is that now self- and antiselfdual parts of
area tensors and of the rotation matrices are, first, complex, second, are related to
each other by complex conjugation. Therefore a more careful analysis is required.

Note that integration over the invariant measure is representable as
\begin{equation}                                                                
\label{Haar}%
\int{f(R){\cal D}R} = \int{f(m)\delta^{10}(\eta^{ab}m_{af}m_{bg}-\eta_{fg}){\rm
d}^{16}m},
\end{equation}

\noindent where $\eta_{ab}$ = ${\rm diag}(\pm 1,1,1,1)$ in the Euclidean/Lorentzian
case. Invariance of this measure is evident\footnote{In fact, originally in our
three-dimensional analysis \cite{Kha2} the invariant measure just arises in the form
analogous to (\ref{Haar}) upon introducing, as in \cite{Wael}, the variables $P_{ab}$
= $l^c\Omega_a^{~f}\epsilon_{cfb}$ and $\Omega^{ab}$ for each edge which are
canonically conjugate in the usual sense. Then we get $\delta$-functions in the
functional integral which take into account the II class constraints on $P$, $\Omega$,
namely, $\delta^6(\bar{\Omega}\Omega-1)\delta^6(\bar{\Omega}P+\bar{P}\Omega)$. The
invariant measure $D\bl{\cal D}\Omega$ follows upon integrating these
$\delta$-functions}. Representing $\delta$-function in (\ref{Haar}) as the Fourier
trans\-form over a 10-component variable symmetrical matrix $\lambda^{ab}$ we can
easily per\-form Gaussian integration over $m_{ab}$. As a result, we get
\begin{eqnarray}                                                                
\int{e^{\textstyle i\pi\circ R}}{\cal D}R & = & \int{\exp{\left [{i\over
2}\pi^{ab}m_{ab}+i\lambda^{ab}(\eta^{fg}m_{af}m_{bg}-\eta_{ab})\right ]}{\rm
d}^{16}m_{ab}{\rm d}^{10}\lambda^{ab}}\nonumber\\& = & \int{{{\rm
d}^{10}\lambda^{ab}\over (\det{\lambda^{ab}})^2}\exp{\left [-{i\over
16}\pi^{fa}\pi^{gb}\eta_{fg}(\lambda^{-1})_{ab}-i\eta_{ab}\lambda^{ab}\right ]}}.
\end{eqnarray}

\noindent This is in some sense a matrix analog of a Bessel function. Let us make the
formal substitution of the variables $\lambda^{ab}$ =
$\sum\limits_{f,g}{(\eta^{-1/2})^{af}(\eta^{-1/2})^{bg}\Lambda^{fg}}$ where for
definiteness $(\eta^{\pm 1/2})^{ab}$ = ${\rm diag}(\pm i,1,1,1)$ for the Lorentzian
case. The result reads
\begin{eqnarray}                                                                
\label{Lorentz_measure}%
\hspace{-10mm}\int{e^{\textstyle i\pi\circ R}}{\cal D}R =&&\nonumber\\ & &
\hspace{-40mm}\int{{{\rm d}^{10}\Lambda^{ab}\over(\det{\Lambda^{ab}})^2}
\exp{\left[-{i\over16}\sum_{c,d}\pi^{fc}\pi^{gd}(\eta^{1/2})^{ca}
(\eta^{1/2})^{db}\eta_{fg}(\Lambda^{-1})_{ab}-i\delta_{ab}\Lambda^{ab}\right]}}.
\end{eqnarray}

\noindent Real $\lambda^{ab}$ imply purely imaginary $\Lambda^{0i}$ ($i$ = 1, 2, 3).
Let us make rotations in the complex plane to the purely real $\Lambda^{0i}$. The
possibility to proceed in such the way is provided by the sufficiently rapidly
decreasing expression under the integral sign at $|\Lambda^{0i}|$ $\rightarrow$
$\infty$ both separately for each $i$ and simultaneously. Note that the dangerous in
this respect exponential factor $\exp{(-i\delta_{ab}\Lambda^{ab})}$ does not contain
$\Lambda^{0i}$ while $\det{\Lambda^{ab}}$ is bilinear in each $\Lambda^{0i}$ at almost
all the values of the rest of variables. As a result, integrations over the circles of
a large radius $|\Lambda^{0i}|$ = $const$ $\rightarrow$ $\infty$ result in zero at
almost all the values of the rest of variables $\Lambda^{ab}$. Therefore
$\Lambda^{ab}$ in (\ref{Lorentz_measure}) can be considered real.

An advantage of the formula (\ref{Lorentz_measure}) is that the same expression can be
found also for the Euclidean case, the transition between these two consisting in
substitution of Lorentzian $\sum\limits_{c,d}\pi^{fc}\pi^{gd}(\eta^{1/2})^{ca}
(\eta^{1/2})^{db}\eta_{fg}$ by the Euclidean $-\pi^{fa}\pi^{~b}_f$ and vice versa
(remind that the exponent is monotonic at genuine real area tensor $\pi$ in the latter
case, not oscillating as at earlier used analytical continu\-a\-ti\-on $\pi$ $\to$
$-i\pi$). One of the conditions imposed when passing from the area tensor Regge
calculus (where $\pi$ are independent tensors) to the ordinary Regge calculus (that is
to say, on physical surface) is $\pi*\pi$ $\equiv$ $\pi^{ab}\!\!$ $\!\!\pi^{cd}$
$\!\!\epsilon_{abcd}/4$ = 0, therefore $\pi^{ab}\pi_{ab}$ is the only scalar which
could be constructed of $\pi^{ab}$. Thereby this is what should be replaced by
$-\pi^{ab}\pi_{ab}$ when passing from the Euclidean metric to the Lorentzian one. Thus
we find the same exponential cut-off factor in the Lorentzian measure as $\nu
(|\bpi|)$ in the Euclidean one,
\begin{equation}                                                               
\label{Euclidean_measure1}%
\int{e^{\textstyle i\pi\circ R}{\cal D}R}\Rightarrow {\nu (|\bpi |)^ 2\over |\bpi
|^4}, ~~~~ \nu (l)={l \over \pi}\int\limits_{0} ^{\pi}{{{\rm
d}\varphi\over\sin^2{\!\varphi}}\,e^{\textstyle -l/\sin{\varphi}}},
\end{equation}

\noindent but with the argument
\begin{equation}                                                                
\label{norma-pi-lorentz}%
|\bpi| = \sqrt{\sum_i{(\pi^{0i}})^2-\sum_{i<j}{(\pi^{ij}})^2}.
\end{equation}

If $\pi^{ab}$ is timelike, i.e. the components $\pi^{0i}$ dominate, we find still
positive measure. According to the definition of $\pi^{ab}$ as dual tensor, this
corresponds to the spacelike (real) area. If the area is imaginary (timelike), the
measure oscillates and is even complex and does not admit the usual probabilistic
interpretation. Physically, this does not seem surprising since not all the areas (or
lengths) {\it must} fluctuate, some of them should be fixed by hand to specify
triangulation (in analogy with gauge lapse-shift vectors in the ordinary continuum
general relativity). At the same time, the set of functions of $\pi^{ab}\pi_{ab}$ on
which the measure (as functional) is positive seems to be even larger than in the
Euclidean case. This set includes, first, positive functions with support on the
(negative) semiaxis of $\pi^{ab}\pi_{ab}$, such as $-\pi^{ab}\pi_{ab}\theta
(-\pi^{ab}\pi_{ab})$ ($\theta$ is Heaviside function) as it follows from the
considered explicit positivity of the measure on this semiaxis. Second, rotation of
contours in the complex plane of $\pi^{ab}$ immediately in the Lorentzian expectation
value $<f(-\pi^{ab}\pi_{ab})>$ reduces this value to the Euclidean
$<f(\pi^{ab}\pi_{ab})>$; therefore the measure is positive also on the {\it
analytical} functions positive on the negative semiaxis of $\pi^{ab}\pi_{ab}$, e. g.
$-\pi^{ab}\pi_{ab}$.

Thus, the quantum measure is positive on the area tensors $\pi^{ab}_{\sigma^2}$
spacelike w.r.t. the local frame indices. On the other hand, the notation for the
triangle $\sigma^2$ in the index of $\pi^{ab}_{\sigma^2}$ can be considered as an
analog of the world index, or, more accurately, a pair of such indices. An analog of
the spacelike components of area tensor w.r.t. the world index are
$\pi^{ab}_{\sigma^2}$ for the leaf/diagonal triangles $\sigma^2$. Now we observe
consistency between the properties of the quantum measure on area tensor w.r.t. the
local indices  and w.r.t. the world indices of this tensor. Indeed, earlier we have
found that integrations over ${\rm d}^6\pi_{\sigma^2}$ are present in the measure not
for all $\sigma^2$ \cite{Kha} (otherwise we could not, e.g., normalize the measure:
integrations of $\exp (i\pi_{\sigma^2}\circ R_{\sigma^2})$ would give the product of
$\delta$-functions of (the antisymmetric part of) $R_{\sigma^2}$ which are not
independent due to the Bianchi identities \cite{Reg}, now for the matrices
$R_{\sigma^2}(\Omega)$). Some regular choice is that ${\rm d}^6\pi_{\sigma^2}$ are
present in the measure only for the leaf and diagonal triangles $\sigma^2$ (or,
rather, this can serve a definition of what do we call the leaf/diagonal triangles).
Correspondingly, only $\pi^{ab}_{\sigma^2}$ for the leaf and diagonal triangles do
fluctuate. It is important that the quantum measure be positive on the sufficiently
large subspace of possible fluctuations of those objects which do fluctuate. This just
takes place: the areas fluctuate which are spacelike in the world index, that is, from
3D subset, and the quantum measure is positive also on the spacelike areas, now w.r.t.
the local frame index, that is, again on 3D subspace (or, taking into account also the
Euclidean case, not smaller than 3D subspace). It looks surprising that consistent are
the things which seem to have perfectly different origins. Namely, properties of the
quantum measure w.r.t. the world index of area tensor are consequence of the structure
of the Bianchi identities on the curvature matrices $R_{\sigma^2}$, whereas considered
in the present paper properties of the quantum measure w.r.t. the local frame index
refer to the dependence of the measure on a single area. The analogous situation had
been encountered also in the three-dimensional model \cite{Kha2}.

\bigskip

The present work was supported in part by the Russian Foundation for Basic Research
through Grant No. 05-02-16627-a.

\end{document}